\documentclass[preprint]{aastex}
\usepackage[dvips]{epsfig}

\shorttitle{Populating Orbits in a Bar Potential}
\shortauthors{Barnes and Tohline}

\begin{document}
\title{Populating Stellar Orbits Inside a Rotating, Gaseous Bar} 
\author{Eric I. Barnes and Joel E. Tohline}
\affil{Department of Physics \& Astronomy, Louisiana State University, Baton Rouge, 
LA 70803}
\email{tohline@physics.lsu.edu}
\email{barnes@physics.lsu.edu}

\begin{abstract}
In an effort to better understand the formation and evolution
of barred galaxies, we have examined the 
properties of equatorial orbits in the effective potential of one specific
model of a rapidly rotating, steady-state gas-dynamical bar that has
been constructed via a self-consistent hydrodynamical simulation.   
At a given value of the Jacobi constant, roughly half of all test 
particles (stars) that are injected into the equatorial plane of this potential 
follow quasi-ergodic orbits;  most regular prograde orbits have an overall 
``bowtie'' shape; and some trace out trajectories that resemble the 
$x_4$ family of regular, retrograde orbits.  The bowtie orbits appear to be 
related to the 4/1 orbit family discussed by Contopoulous (1988),
but particles moving along a bowtie orbit pass very close to the
center of the bar twice each orbit.  Unlike the bar-like configurations
that previously have been constructed using dissipationless, N-body 
simulation techniques, the effective potential of our gas-dynamical bar
is very shallow and generally does not support the $x_1$ family of orbits. 
\\
\indent
If primordial galaxies evolve to a rapidly rotating bar-like configuration 
before a significant amount of star formation has taken place, and then 
stars form from the gas that makes up the bar, the initial stellar distribution 
function should consist of orbits that are (a) supported by the gaseous
bar-like potential and (b) restricted to have initial conditions dictated by
the gas-dynamics of the bar.  With this ``Restriction Hypothesis'' in
mind, we propose that stellar dynamical systems that form from gaseous bars 
will have characteristics that differ significantly from systems that form 
from a bisymmetric instability in an initially axisymmetric stellar system.  
Since bowtie orbits are preferred over $x_1$ orbits, for
example, such systems should have a more boxy or peanut shape when seen face-on; 
there will be a mechanism for funnelling material more directly into the center 
of the galaxy; and, near the galaxy center, stars may appear to move along
retrograde trajectories.
\end{abstract}


\section{Introduction}
\subsection{Background}
\indent
Bars are present in at least half of all disk galaxies, including
our own Milky Way \citep[ \S6.5]{ger99,bnt87}.  Accordingly, there 
have been many observational and theoretical 
studies of bar attributes over the last few decades.  
Rather than attempting to summarize this extensive literature here,
we refer the reader to publications from two recent conferences on
the topic of barred galaxies \citep{sal96,bce96} and to the reviews by 
\citet{ath84}, \citet{bnt87}, \citet{cng89}, and especially \citet{sew93}.
Following the lead of
\citet{conetal89}, we find it useful to group previously published
theoretical studies of the structure and stability of barred galaxies 
into the following four broad categories:
\begin{enumerate}
\item Orbit calculations for various two-dimensional (2D) analytical, bar 
      potentials.
\item N-body simulations of stellar bar formation.
\item N-body simulations of interstellar cloud collisions in bars.
\item Studies of gas-dynamical flows in externally prescribed ``stellar'' bar 
      potentials.
\end{enumerate}
Our present work does not fit naturally into any of these categories
because it involves a detailed analysis of the properties of a 
self-gravitating gas-dynamical, rather than stellar-dynamical, bar-like
configuration.
However, there are strong parallels between our study and the 
analysis presented by \citet{sps87} of a purely 
stellar-dynamical bar, so we will begin by reminding the reader of 
the key features of this earlier work.
\\
\indent
In an effort to ascertain not only what kinds of orbits 
are allowed in the potential well of rapidly rotating, barred 
galaxies but also which orbit families are likely to be
populated by stars in such galaxies, \citet[hereafter SS]{sps87} 
combined the tools and analysis techniques that previously had been 
associated with the separate categories of investigation listed as
items 1 and 2, above.   First, they used a 2D N-body code to construct 
a steady-state model of a rapidly rotating, infinitesimally thin,
bar (initially including a small axisymmetric bulge component, a surrounding 
axisymmetric disk, and a ``hot'' component).   Then, using  a standard 
``shooting technique'' along with
surfaces of section and a characteristic diagram, they mapped out
the properties of available (stable and unstable) orbits in the
effective potential well of this numerically generated, 2D steady-state 
bar.      Finally, they identified
which of the numerous possible orbit families were actually being
populated by particles in their N-body simulation.
Among other things, the SS work provided strong evidence in 
support of a ``restriction hypothesis''\footnote{We have chosen to use 
the phrase ``restriction hypothesis'' here
in an effort to encapsulate the essence of the first sentence of \S 4.3
in SS as well as the similar implication that appears in \citet{tes85}.
This phrase does not appear in either the SS or the \citet{tes85} reference.} 
first alluded to by \citet{tes85},
namely, that a real barred galaxy contains stars 
that largely follow a favorable subset of all possible orbit families.
More specifically, SS found that the stellar distribution function
that was associated with their steady-state bar, 
${\rm DF}_{\rm SS}$, was dominated
by particles whose trajectories were associated with, and generally 
trapped around the $x_1$ family of orbits as defined by \citet{cnp80}.
\\
\indent
This restriction hypothesis 
has received additional support 
from \citet{pff91}, who  extended the SS
analysis to fully three-dimensional (3D) N-body models
of steady-state bars, as well as from the study of \citet{bhsf98},
in which a small amount of gas ($8\%$ of the total galaxy mass)
was included in a self-consistent fashion along with a 3D N-body
simulation.
It is not obvious how a nonlinear dynamical simulation -- 
and by inference a real galaxy -- that starts from a nearly axisymmetric
distribution function, ${\rm DF}_{\rm axisym}$, is able to
preferentially select this restricted set of orbits (primarily related to the
$x_1$ family)
while evolving to a bar-like configuration, but the outcome makes 
sense.  Indeed, other generally available orbits, such as orbits associated
with the $x_2$ or $x_4$ families \citep{cnp80}, have trajectories that 
generally do not support the overall shape of the bar.
\\
\indent
Here we perform an analysis that is very similar to the one presented
by SS, but for a 2D equatorial slice of a 3D steady-state bar that has been 
constructed from a self-gravitating, homentropic, compressible gas cloud 
using a finite-difference hydrodynamic technique 
\citep{cphd99,caz99}.   We have been
motivated to conduct this analysis, in part, because the steady-state
gas-dynamical bars described by \citet{caz99}
are the first detailed models of their kind and we were curious to know
to what degree their global attributes resemble the properties of their 
N-body counterparts.   
By using a shooting technique to inject test particles 
into the potential well of one of the \citet{caz99} bars and then
following the motion of the particles through many orbital periods,
we have been able to produce surface of section diagrams to
facilitate such a comparison.
\subsection{Relevance to the Formation of Galaxies}
\indent
We also have been motivated to conduct this analysis in the
context of studies of the formation of galaxies and the 
earliest generation of stars.   
Models of barred galaxy formation historically have assumed 
that  (i) baryonic material (containing cold gas but no stars) first
settles into a flat, rotationally supported axisymmetric
disk; then (ii) a system of stars forms from this disk with a distribution 
function, such as ${\rm DF}_{\rm axisym}$, that is prescribed by the 
location and motion of the gas from which it formed -- {\it i.e.,}
roughly circular orbits confined to a disk with little
vertical thickness; then (iii) after most of the gas has been converted 
into stars so that the system of stars becomes 
sufficiently self-gravitating, the stellar-dynamical system deforms
into a nonaxisymmetric bar-like configuration because it is 
too cold to remain axisymmetric.  The models presented,
for example, in SS, \citet{pff91}, and \citet{bhsf98} begin at
step ``ii'' in this chronological sequence of events then follow the 
subsequent evolution using entirely (or at least predominantly, in the 
case of Berentzen {\it et al.} 1998) N-body simulation techniques.  
\\
\indent
But another scenario bears consideration.   Just as cold, axisymmetric 
stellar-dynamical configurations are known to be dynamically unstable 
toward a bisymmetric instability if they are sufficiently self-gravitating,
the same is true for fluid configurations.  Based on
the classical analytical studies by Riemann and others of rotating
incompressible fluids, this has been known for over one hundred
years (see Chandrasekhar 1969 for a thorough overview).  More
recently, 3D hydrodynamical techniques have been used
to demonstrate that this same type of bisymmetric instability arises
as well in differentially rotating, compressible fluids.
The eigenfunction that naturally develops as a 
result of the instability has a bar-like structure from which a 
loosely wound, two-armed spiral emerges 
\citep{tdm85,wnt87,pdd96,tipd98}.
That is to say, it exhibits a structure very similar to the one seen
in numerous simulations of purely stellar-dynamical systems that
start from qualitatively similar initial axisymmetric states 
\citep[see the above reviews for additional references]{zh78,ms79,sel80}.
Also like their N-body counterparts, as the
bisymmetric distortion reaches nonlinear amplitude in a gas-dynamical 
system, a significant amount of angular momentum redistribution
can take place in a relatively short amount of time and some mass (the
relatively high specific angular momentum material) is shed into
a roughly axisymmetric equatorial disk  
\citep{dgtb86,wnt88,dyg89,idp00}.
Then a central object
containing a majority of the mass usually settles down into a
bar-like structure that exhibits significant internal streaming
motions but is spinning with a coherent pattern speed about its 
shortest axis.  For two specific models of this
type, \citet{caz99}
have carefully mapped the internal structural and flow properties
of this central bar-like object and have demonstrated 
that it is dynamically robust;
as viewed from a frame rotating with the system pattern speed,
the bar appears to be steady-state and dynamically stable.
\\
\indent
With this in mind, we suggest that a reasonable
alternative to the standard model of barred galaxy formation
is one in which the initially axisymmetric, cold gaseous disk
evolves to a bar-like structure before significant star formation
has taken place, then the stars form from the gas that makes
up the gaseous bar.   Via this scenario, the system of stars that emerges
from the gas will have an initial distribution function,
${\rm DF}_{\rm bar}$, that is
quite different from the ones (such as ${\rm DF}_{\rm axisym}$) 
that have been adopted in various N-body simulations to model
the formation of barred galaxies.  By examining the types of particle 
orbits that are supported by the effective potential of one of
the steady-state bars described by \citet{caz99}, 
we will be in a position to better determine what would be the initial 
distribution function for a system of stars that forms directly from a 
gaseous bar.   In making this determination we will introduce a new 
restriction hypothesis 
that is distinctly different from the one discussed above in the context 
of the \citet{tes85} and SS
papers.   Specifically, after determining via a shooting technique
what orbits are allowed in the potential well of the gaseous bar, we propose 
that the realistic stellar distribution function will not contain all
such orbits but, rather, must be restricted to the subset of 
those orbits that are consistent with the positions and the
velocities of the gas from which the stars would form.  We will illustrate 
to what degree this restriction hypothesis places interesting constraints 
on the resulting stellar distribution function.
\\
\indent
Because it appears that the baryonic component of many, if not most, 
barred galaxies is dominated by stars at the
present epoch, at some point in time most of the gas in the bar must be 
converted into stars before this alternative model of barred galaxy 
formation can be considered plausible.    
It would therefore be interesting to know 
whether or not a stellar dynamical system with a 
distribution function similar to
${\rm DF}_{\rm bar}$ can, by itself, produce a self-consistent bar that 
mimics the original gaseous bar 
configuration.  That is, can the transformation from a barred galaxy that
is predominately gaseous into one that is predominantly stellar be a 
relatively smooth one that essentially preserves the system's overall
geometric shape (and its basic morphological features), or must the 
system relax to an entirely different structure after the majority
of its mass has been converted into stars?  Directly related to this
question is the recent study \citep{lut00} that finds that the fraction
of box and peanut shaped bulges in edge-on galaxies is roughly the same
as the fraction of strongly barred face-on galaxies (see also Bureau \&
Freeman 1999).  The implication is
that the box and peanut bulges are barred galaxies seen in profile.  In
the context of the transformation between gaseous and stellar systems, this
suggests that the relaxation of the emerging stellar system should not radically 
alter the morphology of the initial gaseous bar (if this alternative bar
formation scheme is to remain plausible).
An answer to this
question would also be useful in the context of efforts to understand
the evolutionary connection between galaxies at 
the present epoch and galaxies at high redshift that are 
now being directly imaged \citep[ and references therein]{letal98,detal98,
setal99} and would be particularly relevant to 
observations that are designed to ascertain how barred galaxies
have evolved \citep{abe99,bun99,esk00}.
Furthermore, it would be useful to know whether or not the final 
stellar dynamical configuration that is created via this alternative 
galaxy evolution scenario can in any way be differentiated from the 
stellar bars that are produced via N-body techniques directly from 
${\rm DF}_{\rm axisym}$.   It will be necessary to have an answer to 
this question before we will be able to critically distinguish between 
the standard scenario
of barred galaxy formation and the alternative one being
discussed here. 
We will not attempt to address these followup questions
in the present work, although we expect to do so in the future.
\\
\indent
In what follows (\S 2), we briefly review how \citet{caz99}
constructed two steady-state gas-dynamical bars; summarize the
structural and internal flow properties of the system (their
model B) that we have selected to analyze in detail; then briefly
describe the shooting technique that we have used to probe the properties
of test particle orbits in
the effective potential well of this rapidly rotating, gas-dynamical 
bar.  In \S 3, we use surface of section diagrams and sample particle
orbit trajectories to illustrate the types of stellar orbits that 
can, in principle, be supported inside this bar.  (We do not
consider orbits outside the bar.)  
In an effort to better understand the origin of key orbit asymmetries, in
\S 4 we develop and investigate orbits in analytical potential 
functions that mimic the numerically created bar potential.
In \S 5, we apply our new restriction hypothesis 
in order to ascertain into which of the allowed particle
orbits stars would actually be injected if they formed from the gas that makes
up the bar.  The results of this study are summarized in \S 6. 
\\
\section{Initial Conditions and Tools}
\subsection{The Cazes Bar}
\indent
Self-gravitating, triaxial configurations that either are stationary
in inertial space or are spinning about their shortest axis are of
broad astrophysical interest.   Aside from their relevance to the 
global properties of spiral and elliptical galaxies, spinning triaxial 
configurations are thought to be a stage through which dense cores of 
molecular clouds must evolve in order to produce binary stars 
\citep{leb87,caz99}.  
Such configurations also can arise in the context of the late
stages of stellar evolution \citep{lai93,nct00}.  In recent years interest 
in triaxial compact stellar objects has been renewed because they are 
potentially detectable sources for the gravitational-wave detectors that are 
being constructed worldwide.
\\
\indent
Our theoretical understanding of such structures has grown out of
the general class of incompressible, ellipsoidal
figures of equilibrium originally identified over 100 years ago
by Maclaurin, Jacobi, Dedekind, and Riemann, and recently
studied in detail by Chandrasekhar (1969).  The Riemann S-type ellipsoids,
in particular, are an extremely useful family of equilibrium fluid
configurations because they have analytically prescriptible properties
that span a broad range of geometric parameters.  Unfortunately, Riemann
ellipsoids are not completely satisfactory models of galaxies, 
protostellar clouds, or compact stellar objects because they are 
uniform-density configurations with very simple internal flows, whereas 
most astrophysically interesting systems are centrally condensed objects 
that exhibit a wide assortment of different angular momentum profiles.  
\\
\indent
In an effort to study the rotational fission instability in more realistic
models of protostellar gas clouds, \citet{cphd99}
recently has utilized numerical hydrodynamic 
techniques to construct two different steady-state models of
rapidly rotating, triaxial gas clouds having a compressible (specifically,
an $n = 3/2$ polytropic) equation of state.  These models have been
described in detail by \citet{caz99}.    As far as we have been able
to ascertain, these are the only fully self-consistent models of 
self-gravitating, compressible gas bars with nontrivial internal flows 
that have been presented or discussed in the literature.    Because these 
models provide structures that are more realistic than Riemann ellipsoids, 
we have decided to examine the properties of one of them --- specifically 
the one referred to as ``Model B'' in \citet{caz99} --- here in the context
of the formation and evolution of barred galaxies.  Hereafter we will
refer to this model as the ``Cazes bar.'' 
\\
\indent
The gas-dynamical simulation described by \citet{caz99} that
ultimately produced the Cazes bar
began from a rotationally flattened, axisymmetric, $n=3/2$ polytropic gas 
cloud that was in equilibrium and dynamically stable against axisymmetric
disturbances.  The initial model was constructed with an angular velocity
profile such that, in equatorial projection, the model had uniform
vortensity, where vortensity is defined as the ratio of vorticity to
surface density.  
The model had a ratio of rotational to gravitational potential energy 
$T/|W| = 0.282$ and therefore was sufficiently rapidly rotating that 
it was unstable toward 
the development of a bisymmetric, nonaxisymmetric distortion.
Although primarily bar-like in structure, the eigenfunction of the unstable 
bisymmetric mode had a slight, loosely wound, two-armed spiral character.
Some redistribution of angular momentum occurred via
gravitational torques as the mode grew to nonlinear amplitude.
After approximately 30 dynamical times, the system settled down into
a new, dynamically stable, spinning bar-like structure containing $98\%$ of the
initial cloud mass and $95\%$ of the cloud's original total angular
momentum.  At this point in the system's evolution, Cazes reconfigured
the hydrodynamical code so that the evolution could be continued in
a frame of reference that was rotating at a constant angular frequency,
the pattern frequency of
the bar, and then he followed the system's evolution through an additional 
30 dynamical times.
This extended evolution showed that, to a high degree of accuracy, the 
Cazes bar had settled into a steady-state configuration and
was dynamically stable.
\\
\indent
The overall geometric shape of the Cazes bar is illustrated
well by a map of isodensity contours in the equatorial plane
of the bar, as displayed here in Fig. \ref{phinrho}a.  As detailed in the
last column of Table 3 of Cazes \& Tohline (2000; see also the
bottom panels of their Figs. 8 and 9), the bar extends along the major ($x$)
axis to a dimensionless\footnote{ 
As discussed in \S 3.1 of \citet{caz99}, the 
hydrodynamical simulation that created the ``Model B'' Cazes
bar was performed using a set of dimensionless units so that
the model could be straightforwardly scaled to a variety of
different types of astrophysically interesting systems.  
A so-called ``polytropic'' system of units was adopted in
which $M_0=G=K=1$, where $G$ is the gravitational constant,
$K$ is the polytropic  constant in the ($n=3/2$) polytropic 
equation of state,
and $M_0$ is the total mass of the initial, axisymmetric,
equilibrium configuration from which the Cazes bar formed.
As is tabulated in Table 3 of \citet{caz99}, in 
these units, our Cazes bar has a mass $M = 0.958$, a total
angular momentum $J = 0.941$, a semi-major axis length 
$R_{\rm max} = 8.47$, a pattern frequency $\Omega = 0.522$, 
and a maximum density $\rho_{\rm max} = 6.69 \times 10^{-3}$.
We note that all of the figures in this manuscript show lengths 
that have been additionally scaled to the equatorial radius
($R_{\rm eq} = 7.95$)
of the initial axisymmetric model from which the ``Model B''
Cazes bar was formed; hence, $x_{\rm max} \equiv R_{\rm max}
/ R_{\rm eq} = 1.07$.
The appendix in \citet{wnt87}, for example, shows in 
detail how any physical variable can be converted from this
``polytropic'' system of units to more familiar dimensional units.
By way of illustration, when the Cazes bar is scaled to 
$M_0=10^{10} M_{\odot}$ and $x_{\rm max}=2 {\rm kpc}$, it
has a pattern period $P_{\rm pat} = 2\pi/\Omega \approx
1\times 10^7$ yr and a maximum density
of $\approx 3 \times 10^{-22}$ g ${\rm cm}^{-3}$ (see also Cazes 1999).}
radius of $x_{\rm max} = 1.07$, has an intermediate ($y$)-to-major ($x$) axis 
ratio of approximately $0.52$, and possesses two shallow off-axis density
maxima at $|x| = 0.31$.   The two spiral ``kinks'' that are
immediately apparent in the second and fourth quadrants of the 
isodensity contours of Fig. \ref{phinrho}a identify the location of the
two weak standing shocks that accompany the bar's internal 
flow, as described more fully below.
\\
\indent
As discussed by \citet{caz99}, the bar is spinning about 
its shortest ($z$) axis in a counter-clockwise direction 
with respect to Fig. \ref{phinrho}a, with a well-defined pattern 
frequency,\footnote{At the start of its 
``steady-state'' evolution (time $t = 32 \tau_{\rm dyn}$),
the Cazes bar had a dimensionless pattern frequency of
$\Omega = 0.488$, in accord with the value of the frame
rotation frequency $\Omega_0$ that is listed in Table 3 of
\citet{caz99}.  At the end of their simulation ($t = 59 \tau_{\rm dyn}$),
however, the pattern frequency had shifted slightly, to the value
$\Omega = 0.522$ that we will be using here.}
$\Omega = 0.522$, and exhibits a global
ratio of rotational to gravitational potential energy,
$T/|W| = 0.235$.  The bar appears to be spinning as a solid object but,
in reality, it is not.  Instead, as viewed from a frame spinning with
the bar's pattern frequency, each Lagrangian fluid element
in the bar moves along a well-defined streamline in a
periodic, prograde orbit (counter-clockwise in Fig. \ref{phinrho}a) 
with a frequency that varies with position along the streamline.  
The nested fluid streamlines (see the bottom panel of
Fig. 9 in Cazes \& Tohline 2000) do not cross one another, but 
streamlines associated with the lowest density (outermost)
regions of the bar contain a pair of standing
shock fronts.  The velocity of fluid elements that follow
these outermost streamlines becomes supersonic (in the frame
rotating with the bar) as they ``fall'' along the length
of the bar then, with the aid of the shock, the flow becomes
subsonic in order to bend around the end of the bar.  The
two standing shocks are evidenced by the kinks in the 
isodensity contours displayed in Fig. \ref{phinrho}a; see also the
related violin Mach surface in the bottom panel of Fig. 8 in
\citet{caz99}.  Moving radially outward along the shock, the flow exhibits
Mach numbers that vary smoothly from 1.0 to roughly 2.0 (see the discussion
in \S \ref{shocksec} for more details).  Hence, along its entire length, the
standing shock is relatively weak.
\\
\indent
Figure \ref{phinrho}b shows equipotential contours of the effective potential,
\begin{equation} \label{phie}
\Phi_{\rm eff}(x,y) \equiv \Phi(x,y) - \frac{1}{2}\Omega^2 (x^2 + y^2),
\end{equation}
that is generated in the equatorial plane by the rotating Cazes bar.
Hereafter we will refer to the numerically determined effective potential
of the Cazes bar as $\Phi_{\rm CB}$.
Notice that, as with simpler models of rotating bars or
oval distortions, e.g. \citet[ \S3.3.2]{bnt87}, $\Phi_{\rm CB}$
displays four prominent extrema outside of the central, elongated potential
well.  Two relative maxima appear above and below the bar (these are
associated with the traditional $L4$ and $L5$ Lagrange points), and
two saddle points (associated with the $L1$ and $L2$ Lagrange points)
are marked by asterisks to the left and right of the bar.  The $L1$ and $L2$ 
points are located at a dimensionless distance $R_{L2}=1.36$  
from the origin and, for all practical purposes, define the maximum 
extent of the bar along the major axis.  
The solid curves in  Figs. \ref{compphi}a and \ref{compphi}b show the quantitative variation in 
$\Phi_{\rm CB}$ along the major and intermediate axes, respectively, of the bar.
Along the intermediate axis, for example, the effective potential varies from a value
$\Phi_{\rm min} = -1.018$ at $y=0$ to a value associated with the $L4$ and $L5$ 
maxima of $\Phi_{L4,L5} = -0.503$.   
Along the major axis the effective potential climbs to a somewhat lower value, 
$\Phi_{L1,L2} = -0.603$, before dropping again at positions $|x| > R_{L2}$.
As Fig. \ref{phinrho}a illustrates, the Cazes bar has two mild off-axis density 
maxima.  These density maxima help support a corresponding pair of slight 
off-axis minima in the effective potential.  The minima are not immediately 
evident from the contour levels used in  Fig. \ref{phinrho}b, but they can be seen in 
Fig. \ref{compphi}a.   We note that the equipotential contours do not trace out simple 
quadratic surfaces --- they have, instead, an overall ``peanut'' shape --- 
and the contours exhibit a slight spiral twist.  As we replace the 
numerically generated $\Phi_{\rm CB}$ with an analytical ``fit'' (see \S 4), 
we will attempt to mimic these characteristic features.
\subsection{Numerical Techniques and Analysis} \label{numsec}
\indent
Although the \citet{caz99} simulations were fully 3D,
our investigation will be restricted to an analysis of 2D orbits
that reside in the equatorial plane of the Cazes bar.
(We hope to extend this analysis to 3D orbits in the future.)
We will rely heavily upon surface of section diagrams to characterize
the properties of stellar orbits that are allowed in the Cazes bar potential 
and to provide a means by which the general properties of this potential
can be compared with other analytical and N-body potentials that have been 
examined previously in connection with studies of barred galaxies.
\\
\indent
From the appearance of the surfaces of section, we can discern information 
about the various orbits that are supported by the potential.  
For example, if the surface of section 
for a given particle orbit consists of points that form a so-called invariant 
curve, then that orbit has two isolating integrals of motion and is called 
regular.   Orbits that respect only one integral of motion (the particle's 
specific effective energy; see eq.\ref{jacobi}) create a set of apparently 
disorganized points throughout the allowed phase space and hence do not 
produce invariant curves.  These are referred to as irregular, or ergodic, orbits.  
There also may exist a class of orbits that are quasi-ergodic. 
The surfaces of section for these orbits seem to be 
intermediate between regular and irregular.  While a quasi-ergodic orbit 
does not form an invariant curve surface of section, 
neither will it fill the entire phase space.  There are two classes of quasi-ergodic
orbits, stochastic and semistochastic \citep{gns81}.  Stochastic orbit
surfaces of section exclude regions that regular orbits would occupy but, given 
enough time, will otherwise fill the energetically allowed phase space.  Semistochastic
orbits also avoid regions occupied by regular orbits but are further constrained
to not fill all of the energetically allowed phase space.
\\
\indent
We will generally present and discuss the
behavior of groups of particle orbits that 
all have the same Jacobi constant, or specific effective energy, 
\begin{eqnarray} \label{jacobi}
   \epsilon_J &\equiv& E - \vec{\Omega}\cdot \vec{L} \nonumber \\
    &= &\frac{1}{2}(p_x^2 + p_y^2) - \Omega(xp_y - yp_x) + \Phi \nonumber \\
    &= &\frac{1}{2}(\dot{x}^2 + \dot{y}^2) + \Phi_{\rm eff},
\end{eqnarray}
where the canonical momenta are,
\begin{displaymath} \label{canonx}
  p_x = \dot{x} - \Omega y,
\end{displaymath}
and
\begin{displaymath} \label{canony}
  p_y = \dot{y} + \Omega x,
\end{displaymath}
$E$ is the particle's total energy per unit mass, $\vec{\Omega}$ is the 
angular velocity of the bar, $\vec{L}$ is the angular momentum of the 
particle in the rotating frame, and $\Phi_{\rm eff}$ is defined as in 
eq.(\ref{phie}).
\\
\indent 
In order to follow individual particle orbits inside $\Phi_{\rm CB}$,
we will use a simple Verlet integration scheme \citep{ver67}.  
The advantage of using a Verlet algorithm is that there is
good long term conservation of the Hamiltonian.  For the orbits studied in this
paper, the Hamiltonian of individual orbits is conserved to better than $0.5\%$
and the average value of the Hamiltonian is within $0.1\%$ of the specified
Jacobi constant.  That is, each timestep the value of $1/2 (v_x^2 + v_y^2)
+ \Phi_{\rm eff}(x,y)$ changes by at most $0.5\%$ and over a long integration,
the timestep variations tend to cancel each other out.  In our analysis, the
dynamics of each particle is determined solely by the time-invariant
external potential, $\Phi_{\rm CB}$.
This introduces a first difficulty in studying the Cazes bar.
Since $\Phi_{\rm CB}$ is not defined analytically but, rather, is
specified on an $800 \times 800$ Cartesian grid, 
a finite-differencing scheme must be used to evaluate derivatives
of the potential, that is, the acceleration.    For consistency,
when we analyze analytical potentials (see \S 4), they are evaluated on 
the same size grid.  We use a 5 point finite difference stencil in each
direction to
represent derivatives.  As a check on the error introduced with the
finite differences, the analytically derived gradient of one of our 
analytical potentials has been
compared to the finite difference gradient of the same potential defined on an
$800 \times 800$ grid.  The fractional difference is on the order of $10^{-5}$.
The second difficulty is that particle initial positions are chosen to be on 
grid lines, but as the orbit integration progresses, each particle position moves 
continuously.  When the particle's position does not fall precisely on the
intersection of two grid lines, we evaluate both components of the acceleration 
at the four grid points that surround the particle's position, then linearly
interpolate these to the particle's position.  As a check on the scope of 
this problem, we performed an integration with the nonrotating, 
analytical potential studied by \citet{bns82}.  Our resulting orbits and 
surfaces of section were satisfactorily similar to their published results.  
From this we concluded that the gridding of the potential would not wreak 
havoc with the integrations.
\\
\indent
There is some ambiguity in the literature over the question of 
which surfaces of section to use when characterizing a two-dimensional
potential: 
$(x,p_x)$ or $(y,p_y)$.  For example, \citet{bin82} used $(x,p_x)$ 
while SS and \citet{tes85} examined $(y,p_y)$.  In order to glean
as much information as possible about the orbits in $\Phi_{\rm CB}$, 
we have decided 
to look at both the $(x,p_x)$ and $(y,p_y)$ surfaces of section.  
We will base our primary categorizing criteria on the $(x,p_x)$ 
surfaces of section but, as is illustrated below, 
the $(y,p_y)$ surfaces of section also can convey some important
information, so we use them accordingly.
Each $(x,p_x)$ surface of section is obtained
by plotting the $x$-component of the position and canonical momentum
every time the particle crosses the $x$-axis with $p_y > 0$.  
Alternatively, by plotting the $y$-component of the position and canonical
momentum each time the particle crosses the $y$-axis 
with $p_x < 0$, a $(y,p_y)$ surface of section is created.  
\\
\indent
The shooting technique that we have used to investigate orbits in the 
Cazes bar potential is as follows.  First, a value of $\epsilon_J$ is chosen.   
Second, an initial position $x_i$ is selected along the major axis of the bar.  
At this position, we start with $\dot{x}_i=0.0$.  
From $\epsilon_J$, $x_i$, and $|\dot{x}_i|$,
the corresponding value of $|\dot{y}_i|$ is uniquely determined.  
Then a particle trajectory is integrated with each of the four 
combinations of these initial velocity components: 
$(\dot{x}_i,\dot{y}_i)$,$(-\dot{x}_i,\dot{y}_i)$,$(\dot{x}_i,-\dot{y}_i)$,
$(-\dot{x}_i,-\dot{y}_i)$.  
During each orbit integration, the points for the $(x,p_x)$ 
and $(y,p_y)$ surfaces of section
are calculated.  Without changing the initial position, another value of 
$|\dot{x}_i|$ is chosen; we typically proceed in steps of $0.2$.  A new 
$|\dot{y}_i|$ is thereby determined and the integration is repeated for each 
of the four velocity combinations.  
This cycle continues until the maximum allowed value of $\dot{x}_i$ 
has been reached for that initial position.  At that point, the initial 
position is changed and the entire procedure is repeated.  
On our $800 \times 800$ grid, we began 
this entire cycle by selecting the value of $x_i$ corresponding
to the $20^{th}$ zone from the center,
then moved out along the axis in steps of 20 grid zones
until the energetically limiting position along the major axis was
reached.  As a result, we examined roughly 200 unique
orbits for each selected value of $\epsilon_J$.  Finally, we emphasize that,
throughout our presentation, equatorial-plane coordinate axes are always
oriented such that the $x$-axis coincides with the major axis of the bar,
as in Fig. \ref{phinrho}.
\\
\section{Analysis of the Cazes Bar Potential} \label{scbsec}
\subsection{Composite Surfaces of Section}
\indent
Figure \ref{cazxsos} shows a composite $(x,p_x)$ surface of section diagram for 
six separate regular orbits that arise in the Cazes bar potential when the 
Jacobi constant $\epsilon_J = -0.75$.   
(This value of the Jacobi constant
has been selected for illustrative purposes only; it
has no special significance other than it lies between $\Phi_{\rm min}$ and 
$\Phi_{L1,L2}$).  The contour that corresponds to this value 
of $\epsilon_J$ is identified by the dashed-dotted solid line in  Fig. \ref{irreg}a. 
Additionally, the surface of section diagrams in sections 3 \& 4 also display
zero-velocity curves, i.e. the locus of points in the surface of section
diagram at which the potential equals the value of $\epsilon_J$.
Figure \ref{cazxsos} contains:
\begin{itemize}
\item one elongated region (marked by $\times$ symbols) confined to
      a narrow, short segment of the negative $x$ axis;
\item five disconnected regions (marked by squares) that lie mostly at 
      negative values of $x$ and surround the narrow elongated
      region;
\item four disconnected regions (represented by diamonds) having $|p_x|$ 
      values that are generally larger than that of the five regions 
      marked by squares;
\item three islands identified by two separate, but nested surfaces
      of section (marked by asterisks and triangles);
\item a set of three curves (shown as $+$ symbols) that appear to define a 
      boundary between the three islands and the region of the diagram
      occupied by the other surfaces of section.
\end{itemize}

Figure \ref{cazysos} is a $(y,p_y)$ composite surface of section that complements
the $(x,p_x)$ surface of section shown in Fig. \ref{cazxsos}.  
The orbits that create each of the surfaces of section in Fig. \ref{cazysos} are marked
by the corresponding symbols in Fig. \ref{cazxsos}.  For example, the orbit that 
forms the smallest three-island surface of section in Fig. \ref{cazxsos} (marked with 
triangles) creates the skewed ellipse that is centered on $y \approx -0.05$ 
in Fig. \ref{cazysos}.  
Note that the majority of points that make up the $(y,p_y)$ surfaces of 
section fall at negative values of $y$, suggesting that the orbits from 
which they are derived are retrograde.  As we shall show, the surface of
section marked by crosses is derived from an orbit
which appears to be related to the $x_4$ family of retrograde orbits.  The
surfaces of section marked by squares and diamonds belong to retrograde orbits
with higher order resonances.  However,
the orbits associated with the three islands in Fig. \ref{cazxsos} are, in fact,
prograde.
\\
\indent
One striking feature of all of the surfaces of section that make up 
Figs. \ref{cazxsos} and \ref{cazysos} is the lack of symmetry.  In nonrotating, 
bisymmetric potentials, surfaces of section show reflection symmetry about both 
the $x=0$ ($y=0$) axis and the $p_x=0$ ($p_y=0$) axis.
A variety of such symmetric surfaces of section may be found in 
\citet[ \S 3.3]{bnt87}.  Rotating potentials lose the reflection symmetry about 
the $x=0$ ($y=0$) axis, but generally retain it across the $p_x=0$ 
($p_y=0$) axis.  Some examples of surfaces of section with this symmetry intact
may be seen in SS as well as in \citet{tes85}.  Figures \ref{cazxsos} and \ref{cazysos} 
exhibit the
expected rotational based asymmetry with respect to the $x=0$ ($y=0$) axis,
but they also display a slight asymmetry about the $p_x=0$ $(p_y=0)$ axis.  
As we attempt to develop an analytical approximation to $\Phi_{\rm CB}$ in \S 4, 
we will strive to reproduce the primary features seen in Figs. \ref{cazxsos}
and \ref{cazysos}, including this asymmetry.
\\
\subsection{Individual Orbits}\label{indorbs}
\indent
While the surface of section is a useful tool for categorizing orbits,
the orbits themselves are of primary importance.  We begin with a description
of the regular orbits that have just been identified in connection with
the Cazes bar potential.  The frames
in the left column of  Fig. \ref{cazorbs} isolate individual $(x,p_x)$ surfaces of section 
from the  Fig. \ref{cazxsos} composite diagram, while the frames in the right column of
Fig. \ref{cazorbs} illustrate the $x-y$ orbital trajectories from which each 
corresponding surface of section was derived.
\\
\indent
Figure \ref{cazorbs}f shows the nearly closed orbit that leads to the smallest,
three-island surface of section (marked by triangles) illustrated in Fig. \ref{cazxsos}.
This orbit, as well as each of the two closely related orbits depicted in 
Fig. \ref{cazorbs}d and Fig. \ref{cazorbs}b, has the shape of a bowtie.  
Hence we will refer to the 
regions of the surface of section diagrams that are occupied by these orbits
--- the three-islands in $(x,p_x)$ and the skewed ellipses near the origin
in $(y,p_y)$ --- as the bowtie regions.
Particles travel on bowtie orbits in a counter-clockwise direction (i.e.,
the overall motion is prograde) and make four radial oscillations before completing 
one full orbit cycle.   Hence, the orbits illustrated in frames $a-f$ of
Fig. \ref{cazorbs} are almost certainly related to the $4/1$ family of orbits discussed
by Contopoulos (1988; see especially his Fig. 1a).
However, during two of the radial oscillations in a bowtie orbit, the particle
passes very close to and, indeed, around the origin in such a way that its
direction of motion formally becomes retrograde.  This is why the $(y,p_y)$
surface of section for these orbits generally resides at negative values of
$y$.      We note as well that the bowtie orbits do not exhibit perfect reflection 
symmetry about the $x$-axis.  For example, the top and bottom sections of the
orbit shown in Fig. \ref{cazorbs}d seem to be tilted with respect to the intermediate ($y$) 
axis; and the bottom of the ``v'' shape that is formed on the top of the orbit 
shown in Fig. \ref{cazorbs}b does not lie directly above the inverted ``v'' that is formed 
on the bottom of that orbit.  
\\
\indent
The relatively simple orbit shown in Fig. \ref{cazorbs}l is a retrograde orbit.
This is clear from the $(x,p_x)$ surface of section (Fig. \ref{cazorbs}k),
which shows that each time a particle on this orbit crosses the
$x$-axis with a positive $p_y$ it is to the left of the origin
(i.e., at negative $x$), as well as from the
$(y,p_y)$ surface of section (marked with crosses in Fig. \ref{cazysos}),
which shows that each time the particle crosses the $y$-axis
with a negative $p_x$ it is below the origin (i.e., at 
negative $y$).  This orbit is almost certainly a member of
the $x_4$ family of orbits, as defined by \citet{cnp80}.  
\\
\indent
The regular orbits shown in Figs. \ref{cazorbs}h and \ref{cazorbs}j are also
largely retrograde.   However, these orbits are much more 
complex than the one illustrated in Fig. \ref{cazorbs}l.  
Using the terminology of \citet{con88}, Fig. \ref{cazorbs}h displays a $5/1$ orbit;
that is, the orbit makes five radial oscillations for every complete
orbit cycle.  Similarly, Fig. \ref{cazorbs}j displays a $6/1$ orbit.
It is easier to understand why these two nearly closed orbits display,
respectively, four and five disconnected regions in the
$(x,p_x)$ surface of section diagram if, rather than counting radial
oscillations, we count how many y-oscillations the orbit
undergoes before completing one full (horizontal) excursion along
the bar.  In this sense, Fig. \ref{cazorbs}h displays a $4:1$ orbit while
Fig. \ref{cazorbs}j displays a $5:1$ orbit, exactly matching the number of
disconnected regions that arise in the surface of section diagram.
\\
\indent
Like a number of other previously investigated, nonaxisymmetric potentials, the 
Cazes bar potential supports a rich variety 
of regular orbits that have a recognizable $n\!:\!m$ oscillatory pattern, in the 
sense just discussed.  Particles
following these trajectories complete $n$ oscillations perpendicular to the 
major ($x$) axis in the time that it takes them to complete $m$ circuits 
along the major axis.   In the case of a closed orbit in which the oscillations
perpendicular to the major axis actually cross the major axis, such an
orbit would be represented by $n$ distinct points in a $(x,p_x)$ surface of section 
diagram.  However, it is also possible that not every oscillation will cross the
major axis.  As an example of this variety, 
Fig. \ref{15to5}a illustrates a nearly closed, regular  $15\!:\!5$ orbit.  However,
the $(x,p_x)$ surface of section diagram in Fig. \ref{15to5}b displays only 13 islands.
The difference between what is expected (15 islands) and what is observed (13 islands)
is due to the interesting behavior of this particular orbit.  When the two apparently
straight sections of the orbit (at $y\approx 0.2$ and $y\approx -0.2$) are closely
scrutinized, they each definitely exhibit a small $y$ oscillation.  Since neither
of these nearly horizontal segments crosses the $x$-axis, neither generates a
corresponding island in Fig. \ref{15to5}b.
\\
\indent
The Cazes bar potential also allows quasi-ergodic orbits to develop.  
In fact, approximately 40\% of the $\approx 200$ orbits studied at this Jacobi constant 
are quasi-ergodic (as determined from surface of section diagrams).  As mentioned earlier, 
quasi-ergodic orbits wander through 
the bar without an overall shape.  For this reason, they are difficult to discuss 
individually, but collectively they have characteristics of interest.  These 
orbits cross the major axis of the bar many times as they move along the length 
of the bar.  Most importantly, they support the shape of the bar.  A sample 
quasi-ergodic orbit is shown superimposed on equipotential contours of the 
Cazes bar in  Fig. \ref{irreg}a.  The corresponding surface of section is shown in
Fig. \ref{irreg}b.  
\\
\subsection{Composite Surfaces of Section for Varying $\epsilon_J$}
\indent
While the previous sections have dealt with orbits at a single energy, it is interesting
to see phase space structure at various energy (Jacobi constant) levels.  Figure
\ref{ejxsos} contains composite $(x,p_x)$ surface of section diagrams for four separate
values of $\epsilon_J$: (a) $\epsilon_J=-0.96$ (near the bottom of the potential well); (b) 
$\epsilon_J=-0.85$; (c) $\epsilon_j=-0.75$ (this is the same as Fig. \ref{cazxsos}); and 
(d) $\epsilon_J=-0.63$ (almost at the $L1,L2$ energy level).  Figure \ref{ejysos} shows
the corresponding $(y,p_y)$ surfaces of section.  As before, quasi-ergodic surfaces of
section are not shown, but do exist at each of these energies.  The most striking aspect
of these diagrams is their similarity to one another.  The bowtie and $5\!:\!1$ orbits
appear in all diagrams.  One difference among these diagrams is the small loop that
appears at $(x\approx -0.1, p_x\approx 0.0)$ in Figs. \ref{ejxsos}c and \ref{ejxsos}d.
As mentioned in \S3.2, these loops are created by $x_4$ orbits.  More difficult to 
distinguish is the presence of an $x_1$ surface of section in Fig. \ref{ejxsos}d.  This
surface of section is composed of two pieces that lie close to the zero velocity curve 
(outer dotted line).  A better view of the $x_1$ surface of section is shown in Fig. 
\ref{ejysos}d; it is the small loop located at $(y\approx 0.4, p_y=0.0)$.  The
relationship between these ($x_1$, $x_4$, and bowtie) orbits and the energy range over
which they exist is best illustrated by a characteristic diagram, as shown here in 
Fig. \ref{char}.  This diagram displays the location at which each periodic orbit 
crosses the $y$-axis as a function of
the energy (Jacobi constant in this case) of that orbit.  Figure \ref{char} 
demonstrates that the bowtie orbits are the dominant regular orbital family in the 
Cazes bar potential.
\\
\section{Analytical Potentials} \label{apot}
\subsection{Rotating Bar}
\indent
In an effort to better understand why the Cazes bar potential supports this particular 
variety of particle orbits, we have developed an analytical 
potential which shares its major structural features.   The effective potential 
that we have developed empirically has the form,
\begin{equation} \label{aphie}
\Phi_{\rm eff}(x,y) =  N\bigg\{ 1- \left( 1 + \left( \frac{x}{R_{L2}} \right) ^{\alpha} + 
  \left( \frac{y}{qR_{L2}} \right) ^2 \right)^{- n/2} \bigg\}
  -\frac{1}{2}\Omega^2(x^2 + y^2) + \Phi_{\rm min} \, ,
\end{equation}
where $N$ is a normalization factor; $q$ determines the strength of the bar-like
distortion; 
$\alpha$ and $n$ are exponents whose values are to be determined;
and $R_{L2}$, $\Omega$, and $\Phi_{\rm min}$ have the 
same definitions as in the Cazes bar potential.  Note that unlike $\Phi_{\rm CB}$, the
$\Phi_{\rm eff}$ given in eq. (\ref{aphie}) is four-fold symmetric, i.e.,  the potential
looks the same under the transformations $x \to -x$ and $y \to -y$. 
The form of eq.(\ref{aphie}) is certainly not a unique way to model the Cazes 
bar potential.  It has been adopted simply because it produces a potential 
sufficiently similar to the Cazes bar potential as well as supporting orbits
like those discussed in \S \ref{indorbs}.
\\
\indent
After setting $R_{L2} = 1.36$, $\Phi_{\rm min} = -1.018$, and
$\Omega = 0.522$, as in the Cazes bar, we have found that eq.(\ref{aphie}) is a
good fit to $\Phi_{\rm CB}$ if we select the following parameter values:
$q = 0.8$, $N = 0.7$, $\alpha = 4$, and $n = 8$.
Figure \ref{arphi} is a plot of the equipotential contours generated by eq.({\ref{aphie})
with this set of parameters, and the dashed curves in Figs. \ref{compphi}a and 
\ref{compphi}b show
the variation of this analytical potential along its major ($x$) and
minor ($y$) axes.
As Fig. \ref{compphi} demonstrates, along the principal axes of the bar,
this potential matches $\Phi_{\rm CB}$ extremely well everywhere inside 
$R_{L1,L2}$ and $R_{L4,L5}$.  
A comparison between Fig. \ref{arphi} and Fig. \ref{phinrho}b shows, furthermore,
that this analytical function compares favorably throughout the ($x,y$) plane,
although along the diagonals it is somewhat more box-like than $\Phi_{\rm CB}$.
\\
\indent
We have generated numerous surfaces of section for this analytically prescribed effective 
potential 
from orbits with $\epsilon_J=-0.75$.  In this potential, approximately 50\% of the 
orbits studied are quasi-ergodic.  The composite $(x,p_x)$ surface of section 
for some of the regular orbits is shown in Fig. \ref{arxsos};  Fig. \ref{arysos} shows the 
corresponding composite $(y,p_y)$ surface of section diagram.
\\
\indent
Figures \ref{arxsos} and \ref{arysos} closely resemble Figs. \ref{cazxsos} and 
\ref{cazysos}, respectively.  We are therefore confident that the analytical potential 
that has been used to generate Figs. \ref{arxsos} and \ref{arysos} is indeed an 
appropriate model for the Cazes bar. Figures \ref{arorb}a-j show individual 
surfaces of section from the Fig. \ref{arxsos} composite diagram, along with the 
orbits that created them.  
The orbits shown in Figs. \ref{arorb}b, \ref{arorb}d, and \ref{arorb}f are very 
reminiscent of 
the orbits shown in Figs. \ref{cazorbs}b, \ref{cazorbs}d, and \ref{cazorbs}f, 
respectively.  (The orbit equivalent to the one shown in Fig. \ref{cazorbs}h is
not pictured.)  Note, however, that the bowtie orbits now 
exhibit a reflection symmetry about the $y=0$ axis.
Also, we were unable to find $x_4$ orbits in the rotating, 
analytical potential at this energy.  The small lobe marked by $\times$ symbols
in Figs. \ref{arxsos} and \ref{arorb}i corresponds to a retrograde orbit that appears to 
be trapped in the shallow, off-axis minimum that sits on the positive $x$ 
axis (see Fig. \ref{compphi}a). 
\\
\indent
Figure \ref{arorb}h shows a $5\!:\!1$ orbit (in the $y:x$ oscillation notation 
introduced in \S \ref{indorbs}) that resembles the $5\!:\!1$ orbit 
that we found in the Cazes bar (Fig. \ref{cazorbs}j).  The major difference between 
Fig. \ref{arorb}g (\ref{arorb}h) and Fig. \ref{cazorbs}i (\ref{cazorbs}j) is the reflection 
symmetry exhibited by the former and the lack of symmetry in the latter.  However,
the fact that these orbits correspond in overall appearance with those in 
Fig. \ref{cazorbs} is further evidence (along with the similar appearances of Figs.
\ref{phinrho}b and \ref{arphi} and the major and minor axis fits shown in Fig. \ref{compphi}) 
that the analytical potential closely matches the Cazes bar potential.
\\
\subsection{Twisted, Rotating Bar}
\indent
In an effort to construct an analytical effective potential that supports
orbits having all of the asymmetries seen in the Cazes bar orbits,
we have added a slight spiral twist to the potential function given in
eq.(\ref{aphie}).   
Specifically, our chosen rotating, twisted potential has the form, 
\begin{equation} \label{phiet}
\Phi_{\rm eff}(x',y') = N\bigg\{ 1- \left( 1 + \left( \frac{x'}{R_c} \right)^{\alpha} + 
  \left( \frac{y'}{qR_c} \right) ^2 \right)^{- n/2}\bigg\} 
 -\frac{1}{2}\Omega^2(x'^2 + y'^2) + \Phi_{\rm min},
\end{equation}
where 
\begin{displaymath} \label{twistx}
  x' \equiv x\cos(a\sqrt{x^2+y^2}) - y\sin(a\sqrt{x^2+y^2})
\end{displaymath}
and
\begin{displaymath} \label{twisty}
  y' \equiv  x\sin(a\sqrt{x^2+y^2}) + y\cos(a\sqrt{x^2+y^2}).
\end{displaymath}
For this potential, shown in Fig. \ref{tarphi}, $a=0.1$; otherwise the values of the 
parameters are the same as for the rotating effective potential discussed in 
\S 4.1.
\\  
\indent
The composite $(x,p_x)$ surface of section for regular orbits with 
$\epsilon_J=-0.75$ that are supported by this twisted potential is shown 
in Fig. \ref{tarxsos};  Fig. \ref{tarysos} shows the corresponding composite 
$(y,p_y)$ surface of section.
About 50\% of orbits studied in this potential are quasi-ergodic.  Notice 
that the overall composite surface of section bears a strong resemblance to 
that of the untwisted potential.   The major difference lies in the symmetry 
of the surface of section.  The reflection symmetry about the $p_x=0$ 
(and $p_y=0$) axis is now gone.  
An example of this feature is that the 3 islands (marked by triangles and asterisks) 
that are positioned symmetrically in the Fig. \ref{arxsos} surface of section diagram 
are twisted slightly from those positions in Fig. \ref{tarxsos}. 
The effect of the twisting of the potential on the orbits can be seen in Figs. 
\ref{tarorbs}a-j.  The orbits in Fig. \ref{tarorbs} resemble those from \S 3.2 even more 
closely than the orbits shown in Fig. \ref{arorb} in that there is now an asymmetry due to 
the twisting.  
The effective potential given by eq. (\ref{phiet}) appears to provide an excellent
approximation to $\Phi_{\rm CB}$.
\\
\section{Restriction Hypothesis} \label{restrict}
\indent
Up to this point, we have identified many stellar orbits that,
in principle, could be supported by the Cazes bar potential.  
In the context of our Restriction Hypothesis (hereafter RH), we next ask,
``Which of these orbits would be populated by stars that form from the gas 
and, therefore, have initial velocities determined by the gas in the bar?''   
\subsection{Restriction Hypothesis Orbits}
\indent
In order to maintain consistency between the discussion of our RH 
orbits and the previously discussed orbits, we want to focus on the orbits
of stars that are created with a Jacobi constant $\epsilon_J=-0.75$.  
However, we must abandon the method of choosing initial conditions as outlined 
in \S 2.2.  From the gas motions
that are an integral part of the Cazes bar structure, we now have specific 
values of the velocity associated with each coordinate position in the bar.  
With this in mind, Fig. \ref{rhinit} shows contours of constant $\epsilon_J$, where the 
known velocity of the gas has been used in the determination of $\epsilon_J$ 
at each $(x,y)$ location.  The dashed-dotted contour underlying the large assortment of
symbols identifies at what locations in the bar stars could 
form with $\epsilon_J = -0.75$.
Because these contours are dependent on the velocity field of the gas,  
the shocks mentioned in \S 2.1 become noticeable in Fig. \ref{rhinit}, whereas they
were not readily identifiable in our earlier plots of
$\Phi_{\rm CB}$ (Fig. \ref{phinrho}a) or $\Phi_{\rm eff}$ (Figs. \ref{arphi} and 
\ref{tarphi}). 
\\
\indent
In order to investigate the behavior of an assortment of our RH orbits, 
we positioned 30 particles along the $\epsilon_J=-0.75$ contour, as 
shown in Fig. \ref{rhinit}, and assigned to each the velocity of the Cazes bar 
gas at that location.  
For discussion purposes, we divided these particles into two groups: 
one that begins on the positive side of the major axis and terminates
where the $\epsilon_J = -0.75$ contour crosses the intermediate ($y$)
axis; and one that begins near the intermediate axis and ends where
this energy contour crosses the negative side of the major axis.  
The composite $(x,p_x)$ surface of section diagrams that we derived 
from the first and second groups are
shown in Figs. \ref{rh1xsos} and \ref{rh2xsos}, respectively.  Corresponding $(y,p_y)$ 
composite surface of section diagrams for the first and second groups are 
displayed in Figs. \ref{rh1ysos} and \ref{rh2ysos}, respectively.  
The symbols marking the initial positions of the particles in Fig. \ref{rhinit} 
are the same symbols used to make the corresponding surfaces of section in Figs. 
17 and 18.
\\
\indent
By comparing these new figures with Fig. 3,
we are able to assess the
general impact of our RH.  One of the most striking 
differences between these figures results simply from
the fact that we have elected to include the surfaces of section of 
quasi-ergodic orbits in Figs. 17 and 18, whereas the equivalent 
orbits were not displayed in Fig. 3.  
In this way it is clear at a glance 
that the RH permits a mixture of both regular and quasi-ergodic orbits to
be populated.
In particular, 9 of the 30 starting positions shown in Fig. \ref{rhinit}, that is,
$30\%$ of our RH particles, produced quasi-ergodic orbits.
It is also clear that the bowtie region of phase space 
is well-populated within the constraints of our RH.
The holes that appear at the centers of the bowtie orbit ``islands'' 
in Figs. 17 and 18 falsely suggest that a strictly periodic bowtie orbit 
does not arise under our RH.  Instead, these empty regions --- and the analogous 
gaps that appear between some of the other regular orbit surfaces of section --- 
arise because the spacing that we have chosen between initial
particle positions in Fig. \ref{rhinit} was relatively coarse.  With a finer spacing, 
these regions would have been filled by regular bowtie orbits, and we probably
would have identified two points, one for each group, on the precisely periodic 
bowtie orbit.
Most significantly, the composite surface of section diagrams (Figs. \ref{rh1xsos} and
\ref{rh2xsos}) that result from
our RH present a large region of phase space that is completely unoccupied.
This is the region that previously had been occupied by retrograde orbits.
It is obvious, therefore, that under the constraints of our RH,
no true retrograde orbits are produced.  This is perhaps not surprising, given
that all of the gas in the Cazes bar is moving along prograde streamlines.  
\\
\indent
It is informative to study the
sequence of orbits that appears as one moves to different starting
positions along the $\epsilon_J=-0.75$ contour of Fig. \ref{rhinit}, in a 
counter-clockwise fashion starting from the position marked by the plus symbol
on the positive $x$-axis.  This point on the major axis is the initial position
for an orbit that is similar to that shown in Fig. \ref{cazorbs}b.  Moving along the 
contour of constant $\epsilon_J$, each successive initial position gives rise to bowtie 
orbits that are more and more closed.  The fifth and sixth points (a square and 
a $\times$) have nearly identical orbits; they form the innermost three-island 
surfaces of section in Fig. \ref{rh1xsos} and the corresponding innermost curves of the 
bowtie region in Fig. \ref{rh1ysos}.  
A particle starting from a position somewhere between these two points would
probably trace the periodic bowtie orbit.  Proceeding
toward the minor axis, the sequence reverses and the orbits become less closed.  The
twelfth point (marked by a $\times$) is the origin for an orbit that is basically
the same as that for the first point.  The last three initial positions in this group, 
up to and including the point on the minor axis, produce quasi-ergodic orbits.  
These are the orbits that, for example, create the swarm of points 
that surround the three bowtie region islands in Fig. \ref{rh1xsos}.     
\\
\indent
We now discuss the second group of initial positions, whose $(x,p_x)$ surfaces 
of section are displayed in Fig. \ref{rh2xsos} and $(y,p_y)$ surfaces of section are shown
in Fig. \ref{rh2ysos}.  In the absence of 
the shock, it would be reasonable to assume that the progression seen in the first
group would simply be reversed.  However, whereas only three positions
nearest the minor axis gave rise to quasi-ergodic orbits in the first group, 
five positions nearest the minor axis lead to quasi-ergodic orbits in group two.
The sixth position (marked by a $\times$) produces an orbit similar to the
one shown in Fig. \ref{cazorbs}b and leads to a large three-island surface of section in
Fig. \ref{rh2xsos}.  The next four points (ending with the triangle just after the shock)
mirror the sequence in the first group by beginning bowtie orbits that become more 
and more closed.  The eleventh point in the second group (marked by a square) is the 
initial position for the most closed orbit in both groups.  This orbit forms the 
smallest three-island surface of section in Fig. \ref{rh2xsos} and the smallest curve in the 
bowtie region of Fig. \ref{rh2ysos}.  
Continuing towards the major axis, the orbits become less closed.  The
point on the major axis (marked by a diamond) gives rise to an orbit that is the 
same as that orbit associated with the first point among the first group of points. 
Since the number of quasi-ergodic RH orbits is greater for the quadrant containing
a shock, it seems that the presence of a shock (even one as mild as this) can
influence orbital structure. 
We will examine the effects of shocks more directly in the next section.  
\\
\indent
Observations in the solar neighborhood indicate that stars are born with velocities
that have some dispersion about the mean motion of the gas.  With this in mind, we
have examined the properties of stellar orbits that begin with the velocity of the
gas plus a modest random component.  The results
vary little from what has already been presented here under the strict RH,  even if a 
random velocity of up to $30\%$ of the magnitude
of the initial velocity is added.  For larger (but $\le 50\%$) random velocities, orbits 
that were bowtie-shaped with no perturbation maintain their basic morphology, but 
unperturbed quasi-ergodic orbits can be ``kicked'' into regular orbits.
\\
\subsection{Orbits Originating Near Shocks}\label{shocksec}
\indent
If stars were to form from the gas with equal probability at all
locations throughout the Cazes bar, then the distribution function of stars that 
would be created at a Jacobi constant $\epsilon_J = -0.75$ would contain a
uniform mixture of all the orbits discussed in \S5.1.  For example,
most would be quasi-periodic and $30\%$, by number, would be quasi-ergodic.   
In real barred galaxies, however, one usually does not find that star formation 
occurs at a uniform rate throughout the entire volume of the bar.   In particular,
the star formation rate usually is higher in the vicinity of a shock 
\citep[ \S5.1.8]{bnm98}.  Since the Cazes bar model contains shocks, this is an 
ideal opportunity to examine how such a process would impact the resulting
distribution function of newly formed stars.
\\
\indent
In order to model this scenario, we placed four groups of 15 particles in the vicinity
of the shock structure that is evident in the fourth quadrant of 
Fig. \ref{phinrho}a; see Fig. \ref{ishock} for details regarding the distribution of
these particles.  These particles were given the gas velocity corresponding 
to their initial positions, according to our RH.  Note that, as described in \S 2.1,
the shock becomes stronger as the distance from the major axis increases, but along
most of its length, the shock is relatively weak.  More specifically, using the initial
particle positions in Fig. \ref{ishock} as a guide: the diamond located at $y\approx -0.31$
identifies the contour level at which the shock front officially begins (Mach number 1.0);
the flow reaches Mach 1.5 between the square and triangle at $y\approx -0.45$; and at
the lowest density contour shown, the Mach number is approximately 2.0. 
These particles no longer share a common value of $\epsilon_J$.  
Hence, individual surfaces of section for regular orbits are likely to overlap
and it becomes much less useful to produce composite surface of section diagrams.  
For this reason, here we will discuss only individual surfaces of section.  
\\
\indent
From this entire group of 60 initial particle positions, we find that only the three 
particles in the second column and farthest from the $x$-axis (marked by a plus symbol,
asterisk, and diamond) follow quasi-ergodic orbits.  All other particles follow
quasi-periodic orbits.  We focus, then, on this second column of particles.
The particle that began farthest from the $x$-axis (a plus symbol in Fig. \ref{ishock}) 
created the $(x,p_x)$ surface of section shown in Fig. \ref{shkxsos}a, and the corresponding 
orbit shown in Fig. \ref{shkxsos}b.   Moving progressively closer to the 
$x-$axis, most of the particles trace orbits that have the general bowtie shape.  
For example, particles starting from the positions marked by the square
($y_i \approx -0.43$) and the asterisk ($y_i \approx -0.34$) generate the orbits 
shown in Figs. \ref{shkxsos}d and \ref{shkxsos}f.  The orbits shown in Figs. \ref{shkxsos}h 
and \ref{shkxsos}j are followed
by particles that are deep in the central region of the potential well
[initial positions marked by the plus ($\epsilon_J=-0.899$) and diamond 
($\epsilon_J=-0.953$) symbols that appear closest to the
major axis in Fig. \ref{ishock}].  
These orbits are quite thin and have a strong overall bar shape.  
(Note that these orbits do not appear to be thin in the figures because
we have expanded the vertical axis in order to reveal more orbit details.)
\\
\indent  
We believe that the presence of the quasi-ergodic orbits is connected to the 
large velocities that are present in the gas that is located immediately before the
shock.  Since bowtie
orbits have turning points in the vicinity of this shock, stars that are created
with small $x$-velocities in this region (i.e., from the post-shock gas) have a better 
chance to fall onto such an
orbit than do stars that are created with a large $x$-velocity (i.e., from pre-shock
gas).  Basically, these high
velocity stars are shot through the region occupied by bowtie orbits and onto 
the only other available trajectories, that is, quasi-ergodic orbits.  Hence, the 
presence of the shock does influence the trajectories onto which stars will be
injected according to our RH, but in such a way that stars which form from the
post-shock gas are unlikely to end up on quasi-ergodic orbits.
\\
\section{Discussion and Conclusions} \label{concl}
\indent
We have used a standard shooting technique to probe the structure
of a rotating bar-like potential, $\Phi_{\rm CB}$, that arises
from the steady-state gas dynamical bar that was constructed
by \citet[see also Cazes \& Tohline 2000]{cphd99} in a recent
three-dimensional hydrodynamical simulation.   This potential
supports a roughly equal mixture of regular and quasi-ergodic
orbits.  Virtually all of the regular prograde orbits appear to 
belong to a single family that we have described as having a
bowtie shape.  These orbits are almost certainly related to the
$4/1$ family of orbits described by \citet{con88} because
particles on bowtie orbits make four radial oscillations
for each complete azimuthal cycle.   But they differ from the $4/1$
orbit illustrated in, for example, Fig. 1a of \citet{con88}
in that they pass very close to, and around, the center of the potential well
twice each orbit cycle (see, for example, our Fig. 4f).  The Cazes bar
potential also supports a variety of regular retrograde orbits,
including some that appear to be members of the $x_4$ orbit family.
\\
\indent
Our analysis indicates that, over a large range of $\epsilon_J$, $\Phi_{\rm CB}$ 
does not support the family of $x_1$ orbits (see the characteristic diagram, Fig. \ref{char}).  
As illustrated in \S 3, no
such orbits were found with a Jacobi constant of $\epsilon_J = -0.75$;
and after a careful probe at a number of other energy levels, we were only 
able to find a few $x_1$ orbits at energies close to $\Phi_{\rm L2}$.
This is perhaps the most striking difference between $\Phi_{\rm CB}$
and the potential wells that have been generated through self-consistent
N-body simulations.  N-body simulations tend to produce bars with
stellar distribution functions, such as ${\rm DF}_{\rm SS}$, that
are dominated by $x_1$-orbits.
We suspect that this is because the Cazes bar has a higher ratio
of rotational to gravitational potential energy $T/|W|$ than
typical N-body bars and that, along its major axis, the
Cazes bar potential is very shallow.  In order to approximate
this behavior, we were driven to design an analytical effective
potential that, while exhibiting a traditional quadratic dependence
--- {\it i.e.}, changing as $(y/R_{\rm L2})^2$ --- 
along the intermediate axis, changes as 
$(x/R_{\rm L2})^4$ from the center along the major axis.
\\
\indent
We have considered the possibility that galaxies form central
bar-like structures while still in a predominantly gaseous state.
Because it has been constructed in a self-consistent manner, the
Cazes bar presents a reasonable representation of such a newly formed,
gaseous galaxy configuration.
If stars form from the gas in such a barred galaxy, our
proposed Restriction Hypothesis
illustrates the orbits into which the stars would be injected
at the time of their formation.  Our analysis indicates that the
distribution function ${\rm DF}_{\rm bar}$ of such a system of stars 
would contain no retrograde orbits,
but it would consist of a reasonable mixture of quasi-ergodic
orbits and regular prograde orbits predominately related to the bowtie
($4/1$) orbit family.  It is important to emphasize that these stellar
orbits are distinctly different from the orbits that gas particles
follow in the Cazes bar.  Elements of gas are accelerated by local 
pressure gradients as well as by gradients in the underlying gravitational 
potential; also, unlike stellar orbits, gas particle orbits do not 
cross one another.  As illustrated by \citet{caz99}, within
the steady-state Cazes bar the gas moves along closed streamlines that 
are approximately elliptical in shape.
It is safe to say that no stars that form from such a gas flow will have
similarly elliptical orbits.  Searching many different initial conditions
for particles in $\Phi_{\rm CB}$, we were unable to find any orbits that
even approximated the gas streamlines.
\\
\indent
There are two indications from our study that the presence of a shock
front increases the ratio of quasi-ergodic to regular orbits.  First,
in the absence of a shock -- in which case the potential would have
exhibited a four-fold symmetry -- we would have expected the ratio of
quasi-ergodic to regular orbits to have been identical in the two
samples of particles whose trajectories started from the positions
shown in Fig. 16.   As discussed in \S 5.1, however, we found
that more of the test particles in the second group (with
starting positions on or closer to the Cazes bar's second quadrant
shock front) followed quasi-ergodic orbits.  Then, in the case where 
we purposely selected a group of starting positions along the fourth 
quadrant shock (see \S 5.2), we found that particles starting from the 
highest velocity regions of the pre-shock gas landed in quasi-ergodic orbits.  
\\
\indent
There are several interesting points to be made about the bowtie
orbit family and about stars that might be injected into bowtie orbits.   
Although bowtie orbits should certainly be classified as a prograde orbit 
family, stars that move along bowtie orbits will {\it appear} to be 
moving in a retrograde sense on the portions of their orbits that 
are nearest the center of the bar.   Also, any star that moves along
a bowtie orbit will (a) spend most of its time near the ``four corners''
of the orbit and (b) pass very close to the center of the potential
well twice each orbit.  When coupled with
our discovery that a significant fraction of stars that form from gas 
in the Cazes bar will be injected into bowtie orbits, the first of
these points leads us to suggest that a
gaseous bar should produce a ${\rm DF}_{\rm bar}$ that is rather boxy 
or peanut-shaped.  This is in contrast to distribution functions like
${\rm DF}_{\rm SS}$ that are dominated by the $x_1$ family of orbits
and are therefore more elliptical in shape.
The second of these points leads us to suggest
that star formation in a primarily gaseous bar may provide a mechanism
for funneling matter in toward the center of a galaxy in situations where
gas dissipation alone does not work efficiently.  As noted by \citet{nns83},
triaxial potentials can provide a means of transporting stellar mass to a
central black hole.  Stars that travel close to a central black hole can
become tidally disrupted, and the resulting gas can form an accretion disk which
fuels an active galactic nucleus (AGN) (c.f., Evans \& Kochanek 1989; Ho, Filippenko,
\& Sargent 1997).  Admittedly, in our present model we have not examined to
what extent a central point mass will scatter and, thereby, disrupt the regular
bowtie orbit \citep{gnb85}.  However, we find the existence of orbits 
that travel near the center of the potential over such a large range of 
energies ($-0.96<\epsilon_J<-0.63$) intriguing.  We hope to perform an
investigation of this model of AGN fueling in the future.
\\
\indent
We now consider whether a purely stellar-dynamical bar could be created with
a distribution function given by ${\rm DF}_{\rm bar}$.   That is to say, if
a purely gaseous galaxy  were to initially evolve into the form of a
steady-state Cazes bar,  then slowly create stars from the gas, injecting 
them according to our Restriction Hypothesis into the orbits that make up 
${\rm DF}_{\rm bar}$, could a smooth
evolutionary transition be made between the purely gaseous bar and one that
is entirely made up of stars but that otherwise exactly resembles the Cazes bar?
Using a technique similar to Schwarzschild's method \citep{sch79}
or that of \citet{cng88}, it is conceivable that the right combination
of bowtie orbits and quasi-ergodic orbits could be assembled to produce a
steady-state stellar dynamical bar.  And this configuration may even closely
resemble the Cazes bar.   (Given that we have found an analytical 
function $\Phi_{\rm eff}$ that closely approximates $\Phi_{\rm CB}$,
it should be relatively straightforward to conduct such a study.)
However it seems unlikely that a system of stars
that forms according to our Restriction Hypothesis from the Cazes bar could 
lead to such a configuration because the specific distribution of gas in the
Cazes bar is unlikely to produce the required proportion of bowtie and
quasi-ergodic orbits.
For example, if in order to create a steady-state stellar bar one needs
$N_{\epsilon_J}$ bowtie orbits with energy $\epsilon_J$, then there must be
the right proportion of gas with energy $\epsilon_J$ at the proper positions
to form stars for these orbits.  With this additional constraint, it seems 
unlikely that there would be a clean transformation between a gaseous and a 
stellar system.  
We suspect, instead, that after more than half of the gas has been converted
into stars, the entire configuration would dynamically relax to a new  
configuration that is dominated by the collective dynamics of the stars.
Since such an evolution would begin from a relatively high $T/|W|$
configuration that contains a large number of stars in bowtie orbits,
it would be interesting to know whether this final state has a more boxy
or peanut shape than the stellar dynamical configurations that have been
created via N-body simulations from initially axisymmetric distribution 
functions.   It may be necessary to answer this question before we are able
to state with any certainty whether barred galaxies form from initially 
axisymmetric (${\rm DF}_{\rm axisym}$) or nonaxisymmetric 
(${\rm DF}_{\rm bar}$) stellar distributions. 

\acknowledgments
The authors would like to thank David Merritt and Dana Browne for insightful 
discussions on the topics presented here.  Thanks also to our referee, R. Fux,
for many helpful and productive suggestions.
This work has been supported, in part, by funding from the US National Science
Foundation through grant AST 99-87344.  EB additionally acknowledges support
from the Louisiana Board of Regents' LEQSF under agreement LEQSF(1996-01)-GF-08 and
through NASA/LaSPACE under grant NGT5-40035.  This work also has been supported, in
part, by grants of high-performance computing time through NPACI machines at 
the San Diego Supercomputer Center.


\newpage

\begin{figure}
\figurenum{1}
\caption{Representative diagrams of the Cazes bar 
   potential-density pair.  
   (a) Equatorial plane isodensity contours of the ``Model B'' Cazes bar;
   (b) equipotential contours in the equatorial plane of the Cazes bar effective
   potential, $\Phi_{\rm CB}$.  
   The dimensionless length scales, $x$ and $y$, that are used here, as well as in
   all other figures throughout this paper, are defined in terms of ``polytropic''
   units, as described in footnote 2.  Asterisks 
   mark the positions of the $L1$ and $L2$ Lagrange points; 
   Also, in (a) note the presence of off-axis density maxima and two spiral ``kinks''
   indicating the presence of shocks.\label{phinrho}}
\end{figure}

\begin{figure}
\figurenum{2}
\caption{Comparison between 
   $\Phi_{\rm CB}$ (solid curves), and our analytically specified,
   untwisted, rotating effective potential (dashed curves) as described 
   in \S4.1.  (a) Comparison along the major ($x$) 
   axis.  (b) Comparison along the intermediate ($y$) axis.  Note the presence of two
   shallow, off-axis minima in (a).\label{compphi}}
\end{figure} 
 
\begin{figure}
\figurenum{3a}
\caption{The $(x,p_x)$ composite surface of section diagram for 6 
   selected regular orbits with $\epsilon_J=-0.75$ that are supported by 
   $\Phi_{\rm CB}$.  The dotted line surrounding the invariant curves is the
   zero velocity curve.\label{cazxsos}}
\end{figure}

\begin{figure}
\figurenum{3b}
\caption{The $(y,p_y)$ composite surface of section diagram for the 
   same 6 orbits represented in Fig. \ref{cazxsos}.  As in Fig. \ref{cazxsos},
   the dotted line is the zero velocity curve.\label{cazysos}}
\end{figure}

\begin{figure}
\figurenum{4}
\caption{Plots of the 6 individual surfaces of section taken from 
   the Fig. \ref{cazxsos} composite diagram and their 
   corresponding orbits.  Surfaces of section are shown on the left,
   (a,c,e,g,i,k); orbits are on the right, (b,d,f,h,j,l).  The symbols used for each 
   surface of section are the same as in Fig. \ref{cazxsos}.\label{cazorbs}}
\end{figure}

\begin{figure}
\figurenum{4}
\caption{continued \label{corbcont}}
\end{figure}

\begin{figure}
\figurenum{5}
\caption{Plots illustrating the behavior of a $15\!:\!5$ orbit that is
   supported in $\Phi_{\rm CB}$. (a)  Orbit showing 15
   vertical oscillations for every 5 horizontal oscillations. (b)  The corresponding
   $(x,p_x)$ surface of section diagram.\label{15to5}}
\end{figure} 

\begin{figure}
\figurenum{6}
\caption{(a) A quasi-ergodic orbit with $\epsilon_J=-0.75$ that is supported 
   by $\Phi_{\rm CB}$ is shown superimposed on equipotential contours of that potential.
   The dashed-dotted contour drawn at $\Phi_{\rm CB}=-0.75$ also serves as a boundary of
   the area inside which this orbit is confined.  As in Fig. \ref{phinrho},  
   asterisks mark the positions of the $L1$ and $L2$ Lagrange points.
   (b) The $(x,p_x)$ surface of section for this orbit.\label{irreg}}
\end{figure}

\begin{figure}
\figurenum{7}
\caption{Composite $(x,p_x)$ surfaces of section for four different values of $\epsilon_J$. 
   (a) $\epsilon_J=-0.96$; this value of the Jacobi constant traps particles near the
   bottom of the potential well.  (b) $\epsilon_J=-0.85$.  (c) $\epsilon_J=-0.75$.
   (d) $\epsilon_J=-0.63$; this value of the Jacobi constant allows particles to
   move throughout the entire bar.  Note the presence of bowtie and $5\!:\!1$ orbital
   surfaces of section in each frame.  Also, $x_4$ orbits appear only in (c) and (d),
   while $x_1$ orbits appear only in (d) (see also Fig. \ref{char}).\label{ejxsos}}
\end{figure}

\begin{figure}
\figurenum{8}
\caption{Composite $(y,p_y)$ surfaces of section for 4 different values of $\epsilon_J$. 
   The energy levels for (a), (b), (c), and (d) are the same as in Fig. \ref{ejxsos}.
   \label{ejysos}}
\end{figure}

\begin{figure}
\figurenum{9}
\caption{A characteristic diagram for three families of orbits in the Cazes bar potential.
   The dashed-dotted line at $-\epsilon_J=0.603$ marks the value of the potential at
   the $L1,L2$ points.  The + symbols represent the position along the $y$-axis where
   $\epsilon_J=\Phi_{\rm CB}$.  Bowtie orbits exist over the entire energy range that
   exists inside the Cazes bar.  Near the bottom of the potential well ($-\epsilon_J
   \approx 0.85$), the periodic bowtie orbits become fully prograde.  The $x_4$ and 
   $x_1$ families exist over only a limited (higher energy) range.\label{char}}
\end{figure}

\begin{figure}
\figurenum{10}
\caption{Equipotential contours of the effective potential that is defined
   analytically by eq.(\ref{aphie}).  The degree to which this analytical function
   matches the numerically prescribed $\Phi_{\rm CB}$ can be judged by comparing
   this figure to Fig. \ref{phinrho}b.  The dashed curves drawn in Figs. \ref{compphi}a
   and \ref{compphi}b show more quantitatively the behavior of this analytical
   function along its $x$ and $y$ principal axes, respectively, in comparison to the
   behavior of $\Phi_{\rm CB}$.\label{arphi}} 
\end{figure}  

\begin{figure}
\figurenum{11a}
\caption{The $(x,p_x)$ composite surface of section diagram for 5
   selected regular orbits with $\epsilon_J=-0.75$ that are supported by the
   rotating analytical potential described in \S4.1.  As before, the dotted line 
   surrounding the invariant curves denotes the zero velocity boundary.  This 
   diagram should be compared with Fig. \ref{cazxsos}.\label{arxsos}}
\end{figure}

\begin{figure}
\figurenum{11b}
\caption{The $(y,p_y)$ composite surface of section diagram for the
   same orbits represented in Fig. \ref{arxsos}.  Again, the dotted line is the
   zero velocity curve.  This figure should be compared with 
   Fig. \ref{cazysos}.\label{arysos}}
\end{figure}

\begin{figure}
\figurenum{12}
\caption{Plots of the 5 individual surfaces of section taken from the 
   Fig. \ref{arxsos} composite diagram and 
   their corresponding orbits.  Surfaces of section are shown on the left, 
   (a,c,e,g,i); orbits are on the right, (b,d,f,h,j).  A comparison between this 
   figure and Fig. \ref{cazorbs} illustrates the degree to which the analytically
   specified effective potential discussed in \S4.1 supports orbits that are like
   the orbits supported by $\Phi_{\rm CB}$.\label{arorb}}
\end{figure}

\begin{figure}
\figurenum{12}
\caption{continued \label{arorbcont}}
\end{figure}

\newpage
Figure captions begin on this page.
\newpage

\begin{figure}
\figurenum{13}
\caption{Equipotential contours of the rotating and twisted analytical 
   potential that is defined by eq.(\ref{phiet}).  A comparison between this diagram
   and the one shown in Fig. \ref{phinrho}b illustrates the degree to which our
   analytical ``fit'' matches $\Phi_{\rm CB}$.\label{tarphi}}
\end{figure}

\begin{figure}
\figurenum{14a}
\caption{The $(x,p_x)$ composite surface of section diagram for 5 
   selected 
   regular orbits with $\epsilon_J=-0.75$ that are supported by the rotating and 
   twisted analytical potential described in \S4.2.  As before, the dotted line
   marks the zero velocity curve.  This diagram should be compared
   with Fig. \ref{cazxsos}.\label{tarxsos}}  
\end{figure}

\begin{figure}
\figurenum{14b}
\caption{The $(y,p_y)$ composite surface of section diagram for the
   same orbits represented in Fig. \ref{tarxsos}.  The dotted line is the
   zero velocity curve.  This figure should be compared with 
   Fig. \ref{cazysos}.\label{tarysos}}
\end{figure}

\begin{figure}
\figurenum{15}
\caption{Plots of the 5 individual surfaces of section taken from the
   Fig. \ref{tarxsos} composite diagram
   and their corresponding orbits.  Surfaces of section are shown on the 
   left, (a,c,e,g,i); orbits are on the right, (b,d,f,h,j).  A careful comparison
   between these plots and the corresponding ones displayed in Fig. \ref{arorb}
   shows that the slight spiral ``twist'' that has been added to eq.(\ref{aphie})
   in order to generate eq.(\ref{phiet}) produces a ``north-south'' asymmetry
   like the one that arises in $\Phi_{\rm CB}$ (see Fig. \ref{cazorbs}).\label{tarorbs}}
\end{figure}

\begin{figure}
\figurenum{15}
\caption{continued \label{tarorbcont}}
\end{figure}

\begin{figure}
\figurenum{16}
\caption{Contours of constant $\epsilon_J$ where, as discussed in \S5.1,
   the value of $\epsilon_J$ at each coordinate position is given by eq.(\ref{jacobi})
   with $\Phi_{\rm eff}(x,y) = \Phi_{\rm CB}(x,y)$ and the velocity components 
   $(\dot{x},\dot{y})$ are specified by the velocity of the gas at each position in
   the steady-state Cazes bar.  The dashed-dotted contour is for $\epsilon_J=-0.75$; the
   spacing between contours is 0.05.  Also shown are the initial positions of 30
   particles representing stars that form from the gas according to our ``Restriction
   Hypothesis'' with $\epsilon_J=-0.75$.\label{rhinit}}
\end{figure}  

\begin{figure}
\figurenum{17a}
\caption{The $(x,p_x)$ composite surface of section diagram that results
   from following the orbits of the ``first'' group of 15 particles in 
   $\Phi_{\rm CB}$, as discussed in \S5.1.  For each ``section'', the
   initial particle position is identified by the corresponding symbol in the first
   quadrant of Fig. \ref{rhinit}; the initial velocity is specified by the Cazes bar
   gas velocity at each location.  This figure should be compared with Fig. 
   \ref{cazxsos}
   keeping in mind that quasi-ergodic orbits have been included here, whereas for 
   clarity
   they were omitted in Fig. \ref{cazxsos}.  Note that none of the retrograde orbits
   shown in Figs. 3 and \ref{cazorbs} are populated under the RH.\label{rh1xsos}} 
\end{figure}

\begin{figure}
\figurenum{17b}
\caption{Complementing Fig. \ref{rh1xsos}, this shows the $(y,p_y)$ 
   composite surface of section diagram generated by the orbits of the group of
   15 particles identified in the first quadrant of Fig. \ref{rhinit}.  (See the
   caption to Fig. \ref{rh1xsos} for relevant details.)  This diagram should be
   compared with Fig. \ref{cazysos}.\label{rh1ysos}} 
\end{figure}

\begin{figure}
\figurenum{18a}
\caption{The same as Fig. \ref{rh1xsos}, but for the ``second'' group of
   15 particles identified in the second quadrant of Fig. \ref{rhinit}.\label{rh2xsos}}
\end{figure}

\begin{figure}
\figurenum{18b}
\caption{The same as Fig. \ref{rh1ysos}, but for the orbits of particles
   identified in the second quadrant of Fig. \ref{rhinit}.\label{rh2ysos}}
\end{figure}

\begin{figure}
\figurenum{19}
\caption{Initial positions of the four groups of 15 particles placed in the vicinity
   of the shock that is present in the fourth quadrant of the Cazes bar, superimposed on 
   isodensity contours showing the fourth quadrant structure of the bar.  The Jacobi
   constants for the particles that are marked by the vertical column containing a
   variety of symbols are as follows (from most negative $y$ 
   to least negative $y$): -0.617, -0.613, -0.631, -0.653, -0.676, -0.699, -0.724, 
   -0.751, -0.782, -0.814, -0.843, -0.870, -0.899, -0.927, -0.953.  
   See Fig. \ref{phinrho}a for a less magnified view of this region.\label{ishock}} 
\end{figure}

\begin{figure}
\figurenum{20}
\caption{Plots of the $(x,p_x)$ surfaces of section and corresponding 
   orbits produced in $\Phi_{\rm CB}$ by 5 of the 15 
   particles whose initial positions are shown in Fig. \ref{ishock}.  Surfaces of 
   section are on the left, (a,c,e,g,i); orbits are on the right, (b,d,f,h,j).
   In each case, the symbol used in the surface of section matches the symbol used
   to mark the corresponding particle's initial position in Fig. \ref{ishock}.  
   As discussed in \S5.2, the initial particle velocity is specified by the Cazes
   bar gas velocity at the particle's initial position, as prescribed by our RH.
   \label{shkxsos}}
\end{figure}  

\begin{figure}
\figurenum{20}
\caption{continued \label{sxsoscont}}
\end{figure}


\begin{thebibliography}{}
  \bibitem[Abraham {\it et al.}(1999)]{abe99}
      Abraham, R. G., Merrifield, M. R., Ellis, R. S., Tanvir, N. R., \& Brinchmann, J.
      1999, \mnras, 308, 569
  \bibitem[Athanassoula(1984)]{ath84} 
      Athanassoula, E. 1984, \physrep, 114, 319
  \bibitem[Athanassoula(1992)]{ath92} 
      Athanassoula, E. 1992, \mnras, 259, 328
  \bibitem[Athanassoula et al.(1983)]{ath83} 
      Athanassoula, E., Bienaym\'{e}, O., Martinet, L., \& Pfenniger, D. 
      1983, \aap, 127, 349
  \bibitem[Berentzen {\it et al.}(1998)]{bhsf98}
      Berentzen, Heller, Shlosman \& Fricke 
      1998, \mnras, 300, 49
  \bibitem[Binney(1982)]{bin82} 
      Binney, J. 1982, \mnras, 201, 1
  \bibitem[Binney \& Merrifield(1998)]{bnm98} 
      Binney, J., \& Merrifield, M. 1998, 
      Galactic Astronomy, (Princeton:  Princeton Univ. Press)
  \bibitem[Binney \& Spergel(1982)]{bns82} 
      Binney, J., \& Spergel, D. 1982, \apj, 252, 308
  \bibitem[Binney \& Tremaine(1987)]{bnt87} 
      Binney, J., \& Tremaine, S. 1987, 
      Galactic Dynamics, (Princeton: Princeton Univ. Press)
  \bibitem[Bunker(1999)]{bun99}
      Bunker, A. 1999, in ASP Conf. Ser. 191, Photometric Redshifts and
      High Redshift Galaxies, ed. R. Weymann, L. Storrie-Lombardi, M. Sawicki,
      \& R. Brunner (San Francisco:ASP)
  \bibitem[Bureau \& Freeman(1999)]{bnf99}
      Bureau, M., \& Freeman, K. 1999, \aj, 118, 126
  \bibitem[Buta, Crocker \& Elmegreen(1996)]{bce96}
      Buta, R., Crocker, D. A., \& Elmegreen, B. G., eds.
      1996, Barred Galaxies, ASP Conf. Ser. 91
  \bibitem[Cazes(1999)]{cphd99} 
      Cazes, J. 1999, Ph.D. thesis, Louisiana State University
  \bibitem[Cazes \& Tohline(2000)]{caz99} 
      Cazes, J. E., \& Tohline, J. E. 2000, \apj, 532, 1051
  \bibitem[Chandrasekhar(1969)]{chandr69}
      Chandrasekhar, S. 1969, Ellipsoidal Figures of Equilibrium
      (New Haven: Yale Univ. Press)
  \bibitem[Contopoulos(1980)]{con80} 
      Contopoulos, G., 1980, \aap, 81, 198
  \bibitem[Contopoulos(1981)]{con81}
      Contopoulos, G., 1981, \aap, 102, 265
  \bibitem[Contopoulos(1988)]{con88}
      Contopoulos, G., 1988, \aap, 201, 44
  \bibitem[Contopoulos \& Grosb{\o}l(1988)]{cng88}
      Contopoulos, G., \& Grosb{\o}l, P. 1988, \aap, 197, 83
  \bibitem[Contopoulos \& Grosb{\o}l(1989)]{cng89} 
      Contopoulos, G., \& Grosb{\o}l, P. 1989, \aapr, 1, 261
  \bibitem[Contopoulos \& Papayannopoulos(1980)]{cnp80} 
      Contopoulos, G., \& Papayannopoulos, Th. 1980, \aap, 92, 33
  \bibitem[Contopoulos et al.(1989)]{conetal89} 
      Contopoulos, G., Gottesman, S., Hunter, J., \& England, M., 
      1989, \apj, 343, 608
  \bibitem[Driver {\it et al.}(1998)]{detal98}
      Driver, S. P., Fernandez-Soto, A., Couch, W. J., Odewahn, S. C., Windhorst, R. A.,
      Phillips, S., Lanzetta, K., \& Yahil, A. 1998, \apj, 364, 415
  \bibitem[Durisen et al.(1986)]{dgtb86}
      Durisen, R. H., Gingold, R. A., Tohline, J. E., \& Boss, A. P. 
      1986, \apj, 305, 281
  \bibitem[Durisen, Yang, \& Grabhorn(1989)]{dyg89}
      Durisen, R. H., Yang, S., \& Grabhorn, R. 1989, Highlights Astron., 8, 133
  \bibitem[Eskridge {\it et al.}(2000)]{esk00}
      Eskridge, P. B., Frogel, J. A., Pogge, R. W., Quillen, A. C., Davies, R. L.,
      DePoy, D. L., Houdashelt, M. L., Kuchinski, L. E., Ramirez, S. V., Sellgren, K.,
      Terndrup, D. M., Tiede, G. P. 2000, \aj, 119, 536
  \bibitem[Evans \& Kochanek(1989)]{enk89}
      Evans, C., \& Kochanek, C. 1989, \apjl, 346, L13
  \bibitem[Gerhard(1999)]{ger99}
      Gerhard, O. 1999, in ASP Conf. Ser. 182, Galaxy Dynamics, ed. D. Merritt, M. Valluri,
      \& J. Sellwood (San Francisco:ASP)
  \bibitem[Gerhard \& Binney(1985)]{gnb85}
      Gerhard, O., \& Binney, J. 1985, \mnras, 216, 467
  \bibitem[Goodman \& Schwarzschild(1981)]{gns81} 
      Goodman, J., \& Schwarzschild, M. 1981, \apj, 245, 1087
  \bibitem[Ho, Filippenko, \& Sargent(1997)]{ho97}
      Ho, L., Filippenko, A., \& Sargent, W. 1997, \apj, 487, 591
  \bibitem[Imamura, Durisen, \& Pickett(2000)]{idp00}
      Imamura, J. N., Durisen, R. H., \& Pickett, B. K. 2000, \apj, 528, 946
  \bibitem[Lai, Rasio, \& Shapiro(1993)]{lai93}
      Lai, D., Rasio, F., \& Shapiro, S. 1993, \apjs, 88, 205
  \bibitem[Larson(1990)]{lar90} 
      Larson, R. 1990, \pasp, 102, 709
  \bibitem[Lebovitz(1987)]{leb87}
      Lebovitz, N. 1987, in Highlights of Astronomy Vol. 8, ed. D. McNally
      (Boston:Kluwer), 129
  \bibitem[Lilly {\it et al.}(1998)]{letal98}
      Lilly, S., et al. 1998, \apj, 500,75
  \bibitem[L\"{u}tticke {\it et al.}(2000)]{lut00}
      L\"{u}tticke, R., Dettmar, R.-J., \& Pohlen, M., 2000, \aap, in press
  \bibitem[Miller \& Smith(1979)]{ms79}
      Miller, R. H., \& Smith, B. F., 1979, \apj, 227, 785
  \bibitem[New, Centrella, \& Tohline(2000)]{nct00}
      New, K., Centrella, J., \& Tohline, J. 2000, \prd, 62, 064019
  \bibitem[Norman \& Silk(1983)]{nns83}
      Norman, C., \& Silk, J. 1983, \apj, 266, 502
  \bibitem[Petrou(1984)]{pet84} 
      Petrou, M. 1984, \mnras, 211, 283
  \bibitem[Pfenniger \& Friedli(1991)]{pff91}
      Pfenniger, D. \& Friedli, D. 1991, \aap, 252, 75
  \bibitem[Pickett, Durisen, \& Davis (1996)]{pdd96}
      Pickett, B. K., Durisen, R. H., \& Davis, G. A. 1996, \apj, 458, 714
  \bibitem[Sandqvist \& Lindblad(1996)]{sal96}
      Sandqvist, A. \& Lindblad, P. O., eds. 1996,
      Barred Galaxies and Circumnuclear Activity (New York: Springer-Verlag)
  \bibitem[Schwarzschild(1979)]{sch79} 
      Schwarzschild, M. 1979, \apj, 232, 236
  \bibitem[Sellwood(1980)]{sel80}
      Sellwood, J. A. 1980, \aap, 89, 296
  \bibitem[Sellwood \& Wilkinson(1993)]{sew93}
      Sellwood, J. A., \& Wilkinson, A. 1993, Rept. Prog. Phys., 56, 173
  \bibitem[Simard {\it et al.}(1999)]{setal99}
      Simard, L., et al. 1999, \apj, 519, 563
  \bibitem[Sparke \& Sellwood(1987)]{sps87}
      Sparke, L.S. \& Sellwood, J. A. 1987, \mnras, 225, 653
  \bibitem[Teuben \& Sanders(1985)]{tes85}
      Teuben, P. J. \& Sanders, R. H. 1985, \mnras, 212, 257
  \bibitem[Tohline, Durisen, \& McCollough(1985)]{tdm85}
      Tohline, J. E., Durisen, R. H., \& McCollough, M. 1985, \apj, 298, 220
  \bibitem[Toman et al.(1998)]{tipd98}
      Toman, J., Imamura, J. N., Pickett, B. K., \& Durisen, R. H. 1998, \apj, 497, 370
  \bibitem[Verlet (1967)]{ver67}
      Verlet, L. 1967, Phys. Rev., 159, 98
  \bibitem[Williams \& Tohline(1987)]{wnt87} 
      Williams, H. A., \& Tohline, J. E. 1987, \apj, 315, 594
  \bibitem[Williams \& Tohline(1988)]{wnt88}
      Williams, H. A., \& Tohline, J. E. 1988, \apj, 334, 449
  \bibitem[Zang \& Hohl(1978)]{zh78}
      Zang, T. A. \& Hohl, F., 1978, \apj, 226, 521
\end{thebibliography}
\end{document}